# Issues, Challenges and Tools of Clustering Algorithms

**Parul Agarwal[1],M. Afshar Alam[2],Ranjit Biswas[3]**

**[1] Department of Computer Science,Jamia Hamdard,
New Delhi,Delhi-62,India**

**[2] Department of Computer Science,Jamia Hamdard,
New Delhi,Delhi-62,India**

**[3] Manav Rachna International University , Green Fields Colony
Faridabad, Haryana 121001**

## Abstract

Clustering is an unsupervised technique of Data Mining. It means grouping similar objects together and separating the dissimilar ones. Each object in the data set is assigned a class label in the clustering process using a distance measure. This paper has captured the problems that are faced in real when clustering algorithms are implemented .It also considers the most extensively used tools which are readily available and support functions which ease the programming. Once algorithms have been implemented, they also need to be tested for its validity. There exist several validation indexes for testing the performance and accuracy which have also been discussed here.

## 1. Introduction

Clustering is an active topic of research and interest and has its applications in various fields like biology, management, statistics, pattern recognition etc. But, we shall understand its association with data mining. Data mining[3] deals with small as well as  large datasets with large number of attributes and at times thousands of tuples. The major clustering approaches[1,2,4,5]   are Partitional and Hierarchical.The attributes are also broadly divided into numerical and categorical. In Section 2  we give a brief overview of clustering,In section 3, we discuss properties of algorithms ,Section 4  has Challenges of Clustering Algorithms, followed by Validation indexes in

Section 5 and Tools and Softwares of Clustering Algorithms  in Section 6 and Conclusion in Section 7.

## 2. Overview

Though there exist several categories of clustering algorithms,but in this paper we discuss only the partitional and hierarchical approaches.

   The approach is based upon the clustering method chosen for clustering.The clustering methods are broadly divided into Hierarchical and Partitional. Hierarchical clustering performs partitioning sequentially. It works on bottom –up and top-down.The bottom up approach known as agglomerative starts with each object in a separate cluster and continues combining 2 objects based on the similarity measure until they are combined in one big cluster which consists of all objects. .Wheras the top-down approach also known as divisive treats all objects in one big cluster and the large cluster is divided into small clusters until each cluster consists of just a single object. The general approach of  hierarchical clustering is in using  an appropriate metric which measures  distance between 2 tuples and a linkage criteria which specifies the dissimilarity of sets as a function of the pairwise distances of observations in the sets The linkage criteria could be of 3 types [21 ]single linkage ,average linkage and complete linkage.

   In single linkage(also known as nearest neighbour), the distance  between 2 clusters is computed as:

   $D(C_i,C_j)$= min {$D(a,b)$ : where a  Є $C_i$, b Є $C_j$.

   Thus distance between clusters is defined as the distance between the closest  pair of objects, where only  one object from each cluster is  considered.

   i.e.  the distance between two clusters is given by the value of the shortest link between the clusters.  In average Linkage method (or farthest neighbour), Distance between Clusters defined as the distance between the most distant pair of objects, one from each cluster is considered.

In the complete linkage method, $D(C_i, C_j)$ is computed as

$D(C_i, C_j) = Max \{ d(a,b) : a \in C_i, b \in C_j.\}$

the distance between two clusters is given by the value of the longest link between the clusters.

Whereas,in average linkage

$D(C_i, C_j) = \{ d(a,b) / (l1 * l2): a \in C_i, b \in C_j$. And l1 is the cardinality of cluster $C_i$,and l2 is cardinality of Cluster $C_j$.

And $d(a,b)$ is the distance defined.

The hierarchical clustering is represented by n-tree or dendogram.(Gordon 1996). A dendogram depicts how the clusters are related. By cutting the dendrogram at a desired level a clustering of the data items into disjoint groups is obtained.

The partitional clustering on the other hand breaks the data into disjoint clusters or k partitions.These partitions are performed based on certain objective functions like minimizing square error criteria. etc.

Data sets can be found at:
www.kdnuggets.com/datasets/index.html
www.kdd.ics.uci.edu/
www.datasetgenerator.com

## 3. Properties of Clustering algorithms

The clustering algorithms depend on the following properties:

1. Type of attribute handled by algorithm: the various types are ratio,interval based or simple numeric values.these fall in the category of numeric representations.On the other hand,we have nominal and ordinal. An attribute is nominal if it successfully distinguishes between classes but does not have any inherit ranking and cannot be used for any arithemetic. For eg. If color is an attribute and it has 3 values namely red,green,blue then we may assign 1-red,2-green,3-blue.This does not mean that red is given any priority or preference.Another type of attribute is ordinal and it implies ranking but cannot be used for any arithematic calculation.eg. if rank is an attribute in a database,and 1$^{st}$ position say denoted as 1 and second position denoted as 2 in database.

2. Complexity: What is the complexity of algorithm in terms of space and time?

3.Size of database: A few databases may be small but a few may have tuples as high as say thousands or more

4. Ability to find clusters of irregular shape

5. Dependency of algorithm on ordering of tuples in database:Most of the clustering algorithms are highly dependent on the ordering of tuples in database

6. Outlier detection:As defined in [8],outlier detection is a method of finding objects that are extremely dissimilar or

inconsistent with the remaining data.For data analysis applications outliers are considered as noise or error and need to be removed to produce effective results.Many algorithms have been proposed in [9-13] that deal with outlier detection

## 4. Challenges with cluster analysis

The potential problems with cluter analysis that we have identified in our survey are as follows:

1. The identification of distance measure :For numerical attributes, distance measures that can be used are standard equations like eucledian,manhattan, and maximum distance measure.All the three are special cases of Minkowski distance .But identification of measure for categorical attributes is difficult.

2. The number of clusters : Identifying the number of clusters is a difficult task if the number of class labels is not known beforehand.A careful analysis of number of clusters is necessary to produce correct results.Else, it is found that heterogenous tuples may merge or simialy types tuples may be broken into many.This could be catastrophic if the approach used is hierarchical.B'coz in hierarchical approach if a tuples gets wrongly merged in a cluster that action cannot be undone.

While there is no perfect way to determine the number of Clusters, there are some statistics that can be analyzed to help in the process [22-23]. These are the Pseudo-F statistic, the Cubic Clustering Criterion(CCC), and the Approximate Overall R-Squared.

3. Lack of class labels: For real datasets (relational in nature as they have tuples and attributes) the distribution of data has to be done to understand where the class labels are?

4. Structure of database: Real life Data may not always contain clearly identifiable clusters.Also the order in which the tuples are arranged may affect the results when an algorithm is executed if the distance measure used is not perfect.With a structureless data(for eg. Having lots of missing values), even identification of appropriate number of clusters will not yield good results. For eg. missing values can exist for variables,tuples and thirdly,randomly in attributes and tuples.If a record has all values missing,this is removed from dataset.If an attribute has missing values in all tuples then that attribute has to be removed described in [6]. A dataset may also have not much missing values in which case methods have been suggested in [24]. Also, three cluster-based algorithms to deal with missing values have been proposed based on the mean-and-mode method in [24].

5.T ypes of attributes in a database: The databases may not necessarily contain distinctively numerical or categorical attributes.They may also contain other types like nominal,ordinal,binary etc.So these attributes have to be converted to categorical type to make calculations simple.

6.Choosing the initial clusters : For partitonal approach ,we find that most of the algorithms mention k initial clusters to be randomly chosen.A careful and comprehensive study of data is required for the same. Also, if the intial clusters are not properly chosen, then after a few iterations it is found that clusters may even be left empty.Although , a paper in [7] discusses a farthest heuristic based approach for calculation of centers.

# 5. Validation indexes for Clustering Algorithms

Cluster validity measures:
These indexes measure how accurate the results are.It also determines after a clustering algorithm produces its result,how many tuples have been correctly associated with their class labels accurately and how many belong to a class label with which they should not be associated. These indexes can be used to test the performance and accuracy of various algorithms or accuracy of one algorithm for various parameter values like threshold value (if any),or number of clusters e.t.c
Several validity indexes have been proposed till date. We discuss a few of them:

5.1 Sillhouette : this index was proposed by Peter J. Rousseeuw and is available in [13].

Suppose, the tuples have been clustered into k clusters.For each tuple I, let $a(i)$ define the average dissimilarity of i with other tuples in the same cluster.

Then find the average dissimilarity of I with data of another cluster.Continue this for every cluster of which I is not a member.The lowst average dissimilarity of I of any such cluster is represented as $b(i)$

Then,$s(i)=b(i)-a(i)/(max(a(i),b(i))$

If $s(i)$ is close to 1,it means the data has been properly clustered.

5.2  C- index:

This index [17] is defined as follows:

$C= S- Smin/S max - S min$

Where, S is the sum of distances over all pairs of patterns from the same cluster. Let n be the number of those pairs. Then Smin  is the sum of the n smallest distances if all pairs of patterns are considered .And S max  is the sum of the n largest distance out of all pairs. Hence a small value of C indicates a good clustering.

5.3 Jaccard index

In this index[19] the level of agreement between a set of class labels C and a clustering result K is determined by the number of pairs of points assigned to the same cluster in both partitions:

$J(C,K)= a /(a+b+c)$

where a denotes the number of pairs of points with the same label in C and assigned to the same cluster in K, b denotes the number of pairs with the same label, but in different clusters and c denotes the number of pairs in the same cluster, but with different class labels. The index produces avalue that lies between  0 and 1, where a value of 1.0 indicates that C and K are identical.

5.4  Rand index

This index[20] simply measures the number of pairwise agreements between a clustering K and a set of class labels C.It is measured as

$J(C,K)=  (a  +  d)/(a +b + c + d).$

where a denotes the number of pairs of points with the same label in C and assigned to the same cluster in K, b denotes the number of pairs with the same label, but in different clusters, c denotes the number of pairs in the same cluster, but with different class labels and d denotes the number of pairs with a different label in C that were assigned to a different cluster in K. The index produces a result between 0 and 1.

A value of this index equal to 1 means 100% accuracy and a large value indicates high agreement between C and K.

Many others like the dunn index[15] ,Goodman Kruskal [18],  Davies-Bouldin Validity Index [ 16], have also been proposed .

# 6.  Tools and Softwares of Clustering Algorithms

a) Weka:
Weka[25] is a collection of machine learning algorithms for data mining tasks and is capable of developing new machine learning schemes.It can be applied to a dataset directly or can be initiated from our own java code. Weka contains tools for data pre-processing, classification, regression, clustering, association rules, and

visualization.Procedures in Weka are represented as classes and arranged logically in packages. It uses flat text files to describe the data.

Weka has pre-processing tools(also known as filters) for performing discretization,normalization,resampling, attribute selection,transforming, and combining attributes etc.

b) Matlab Statistical Toolbox [26]
The Statistics Toolbox[26] is a collection of tools built on the MATLAB for performing numeric computations. The toolbox supports a wide range of statistical tasks, ranging from random number generation, to curve fitting, to design of experiments and statistical process control. The toolbox provides building-block probability and statistics functions And graphical, interactive tools.

The first category of function support can be called from the command line or from your own applications. We can view the MATLAB code for these functions, change the way any toolbox function works by copying and renaming

The M-file, then modifying your copy and even extend the toolbox by adding our own M-files.

Secondly, the toolbox provides a number of interactive tools that enables us to access functions through a graphical user interface (GUI).

c)Octave: It is a free software similar to Matlab and has details in [27]

d)SPAETH2[28]
It is a collection of Fortran 90 routines for analysing data by grouping them into clusters

e) C++ code for unix by Tapas Kamingo
It[29] is a collection of C++ code for performing k-means based on local search and LLyod's Algorithm.

f) XLMiner
XLMiner is a toolbelt to help you get quickly started on data mining, offering a variety of methods to analyze your data. It has extensive coverage of statistical and machine learning techniques for classification, prediction, affinity analysis and data exploration and reduction.It has a coverage of both partitional [30]and hierarchical[21] methods .

g) DTREG[31]
it is a commercial software for predictive modeling and forecasting offered ,are based on decision trees,SVM,Neural N/W and Gene Expression programs.For clustering,the property page contains options that ask the user for the type of model to be built(eg. K-means) The tool can also build model with a varying number of clusters or fixed number of clusters.we can also specify the minimum number of clusters to be tried.

If the user wishes,then it has options for selecting some restricted number of data rows to be used during the search process. Once the optimal size is found, the final model will be built using all data rows.

It has parameters like cross validate folds, Hold out sample percentage, usage of training data which evaluate the accuracy of the model for each step. It provides standardization and estimation of importance of predictor values We can also select the type of validation which DTREG should use to test the model.

h) Cluster3
Cluster3[32] is an open source clustering software available here contains clustering routines that can be used to analyze gene expression data. Routines for partitional methods like k means,k-medians as well as hierarchical (pairwise simple, complete, average, and centroid linkage) methods are covered.It also includes 2D self-organizing maps .The routines are available in the form of a C clustering library, a module of perl,an extension module to Python, as well as an enhanced version of Cluster, which was originally developed by Michael Eisen of Berkeley Lab. The C clustering library and the associated extension module for Python was released under the Python license. The Perl module was released under the Artistic License. Cluster 3.0 is covered by the original Cluster/TreeView license.

i) CLUTO
CLUTO [33] is a software package for clustering low- and high-dimensional datasets and for analyzing the characteristics of the various clusters. CLUTO is well-suited for clustering data sets arising in many diverse application areas including information retrieval, customer purchasing transactions, web, GIS, science, and biology. CLUTO provides three different classes of clustering algorithms that are based on the partitional, agglomerative clustering , and graphpartitioning methods. An important feature of most of CLUTO's clustering algorithms is that they treat the clustering problem as an optimization process which seeks to maximize or minimize a particular clustering criterion function defined either globally or locally over the entire clustering solution space. CLUTO has a total of seven different criterion functions that can be used to drive both partitional and agglomerative clustering algorithms, that are described and analyzed in [34-35]. The usage of these criterion functions has produced high quality clustering results in high dimensional datasets As far as the agglomerative hierarchical clustering is concerned,CLUTO provides some of the more traditional local criteria (e.g., single-link, complete-link, and UPGMA) as discussed in Section 2. Furthermore, CLUTO provides graph-partitioning-based clustering algorithms that are well-suited for finding clusters that form contiguous regions that span different dimensions of the underlying feature space.

An important aspect of partitional-based criterion-driven clustering algorithms is the method used to optimize this criterion function. CLUTO uses a randomized incremental optimization algorithm that is greedy in nature, has low computational requirements, and has been shown to produce high-quality clustering solutions [35]. CLUTO also provides tools for analyzing the discovered clusters to understand the relations between the objects assigned to each cluster and the relations between the different clusters, and tools for visualizing the discovered clustering solutions. CLUTO also has capabilities that help us to view the relationships between the clusters,tuples, and attributes.Its algorithms have  been optimized for operating on very large datasets both in terms of the number of  tuples as well as the number of attributes This is especially true for CLUTO's algorithms for partitional clustering. These algorithms can quickly cluster datasets with several tens of thousands objects and several thousands of dimensions.

Moreover, since most high-dimensional datasets are very sparse, CLUTO directly takes into account this sparsity and requires memory that is roughly linear on the input size.

CLUTO's distribution consists of both stand-alone programs (vcluster and scluster) for clustering and analyzing these clusters, as well as, a library via which an application program can access directly the various clustering and analysis algorithms implemented in CLUTO.

Its variants are gCLUTO,wCLUTO

j)Clustan:

Clustan[34] is an integrated collection of procedures for performing cluster analysis.It helps in designing software for cluster analysis, data mining, market segmentation, and decision trees.

# 7. Conclusion

In this paper we have covered the properties of Clustering Algorithms. We have also described the problems faced in implementation and those which affect the clustering results. At last we have described some of the software available that can ease the task of implementation.